\documentclass[twocolumn,showpacs,superscriptaddress]{revtex4}
\usepackage[latin1]{inputenc}
\usepackage[T1]{fontenc}
\usepackage{graphicx}
\usepackage[dvips]{epsfig}
\usepackage{amsmath}
\usepackage{amssymb}
\usepackage{color}
\usepackage{dcolumn}
\usepackage{slashbox}

\begin{document}

\title{Momentum signatures for Schwinger pair production in short
  laser pulses with a sub-cycle structure} 
\author{F.~Hebenstreit}
\author{R.~Alkofer}
\affiliation{Institut f\"ur Physik, Karl-Franzens Universit\"at Graz, A-8010 Graz, Austria}
\author{G.~V.~Dunne}
\affiliation{Department of Physics,University of Connecticut, Storrs, CT 06269, USA}
\author{H.~Gies}
\affiliation{Theoretisch-Physikalisches Institut, Friedrich-Schiller Universit\"at Jena, D-07743 Jena, Germany}
\date{\today}

\begin{abstract}
  We investigate electron-positron pair production from vacuum for short laser pulses with sub-cycle structure, in the nonperturbative regime (Schwinger pair production). We use the non-equilibrium quantum kinetic approach, and show that the momentum spectrum of the created electron-positron pairs is extremely sensitive to the sub-cycle dynamics -- depending on the laser frequency $\omega$, the pulse length $\tau$, and the carrier phase $\phi$ -- and shows several distinctive new signatures. This observation could help not only in the design of laser pulses to optimize the experimental signature of Schwinger pair production, but also ultimately lead to new probes of light pulses at extremely short time scales.
\end{abstract}

\maketitle

Electron-positron pair production due to the instability of the quantum electrodynamics (QED) vacuum in an external electric field is a remarkable non-perturbative prediction of QED \cite{Sauter:1931,Heisenberg:1935,Schwinger:1951nm} that has not yet been directly observed. Significant recent advances in laser technology have raised hopes that the required critical field strength of $E_{\rm cr}\sim 10^{16}$ V/cm may soon be within experimental reach \cite{Tajima:1900zz,Ringwald:2001ib,Gordienko:2005zz}, either in optical high-intensity laser facilities such as Vulcan or ELI \cite{ELI} or in X-ray free electron laser (XFEL) facilities \cite{XFEL}. Observation of this elusive phenomenon in the non-perturbative domain would complement the perturbative multi-photon pair production seen at SLAC E144 using nonlinear Compton scattering \cite{Burke:1997ew}. Moreover, it would represent a significant advance in our understanding of non-perturbative phenomena in quantum field theory, with potentially important lessons for related phenomena such as Unruh and Hawking radiation. The Schwinger mechanism has also been used to study various non-perturbative phenomena: e.g., string-breaking in the strong interactions \cite{Casher:1978wy}, pair production in supercritical fields \cite{Rafelski:1976ts}, neutrino production in a fermionic density gradient \cite{Kachelriess:1997cr}, and saturation in heavy-ion collisions \cite{Kharzeev:2006zm}. Since the basic physics is quantum tunneling, the effect is exponentially weak, and so it is important to search for distinctive signs that might facilitate its detection. Here we consider a realistic laser pulse with sub-cycle structure, and find distinctive new signatures in the momentum distribution of the produced pairs. We also explain these signatures by relating the non-equilibrium quantum kinetic approach \cite{Kluger:1991ib,Kluger:1992ib,Rau:1995ea, Kluger:1998bm,Smolyansky:1997fc,Schmidt:1998vi} to the quantum mechanical scattering description \cite{Brezin:1970xf,Popov:1972,Popov:1973az}.

\begin{figure}[bh]
  \centering
  \includegraphics[scale=0.8]{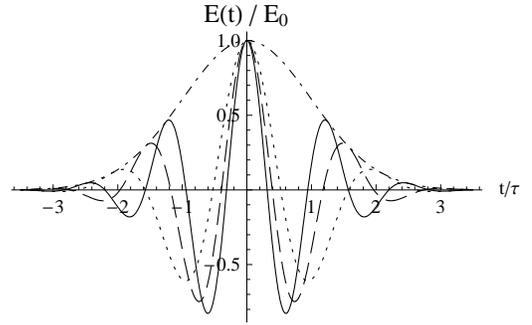}
  \caption{\label{fig:pulseshape} Shape of the electric field
    Eq.~(\ref{eqn:elfield}), for carrier-phase $\phi=0$, when passing from $\sigma=3$ (dotted
    line) to $\sigma=4$ (dashed line) to $\sigma=5$ (solid line). The pure
    Gaussian field (dashed-dotted line) is given as reference.}
\end{figure}

The original estimates \cite{Sauter:1931,Heisenberg:1935,Schwinger:1951nm} assumed a constant and uniform external electric field, but realistic ultra-strong fields are realized in short pulse, focussed lasers. We concentrate here on the time dependence of the electromagnetic field and neglect spatial variations, assuming that the spatial focussing scale is much larger than the Compton wavelength. This approximates the experimental situation of two counter-propagating short laser pulses, generating a standing-wave electric field which is approximately spatially homogeneous in the interaction region, such that $\vec{E}(t)=(0,0,E(t))$ with (see Fig. \ref{fig:pulseshape})
\begin{equation}
  \label{eqn:elfield}
  E(t)=E_0\cos(\omega t+\phi)\exp\left(-\frac{t^2}{2\tau^2}\right)\ .
\end{equation}
Here $\omega$ is the laser frequency, $\tau$ defines the total pulse length, and $\phi$ is the 'carrier phase' (carrier-envelope absolute phase). We are motivated to investigate the carrier-phase dependence of the Schwinger
mechanism by the sensitive carrier-phase dependence of strong-field ionization experiments in atomic, molecular and optical (AMO) physics \cite{Brabec:2000zz}. It is convenient to introduce the parameter $\sigma=\omega\tau$ as a measure of the number of oscillation cycles within the Gaussian envelope pulse.
This type of electromagnetic field configuration can be represented by a time-dependent vector potential $\vec{A}(t)=(0,0,A(t))$:
\begin{equation}
  \label{eqn:vecfield}
  A(t)=-\sqrt{\frac{\pi}{8}}e^{-\sigma^2/2+i\phi}E_0\tau\operatorname{erf}\left(\frac{t}{\sqrt{2}\tau}
-i\frac{\sigma}{\sqrt{2}}\right)
  + c.c.  
\end{equation}
Since Schwinger pair production is a time-dependent nonequilibrium process, quantum kinetic theory provides an appropriate framework \cite{Kluger:1998bm,Schmidt:1998vi,Bloch:1999eu,Alkofer:2001ik,Vinnik:2002qk, Roberts:2002py,Hebenstreit:2008ae,Blaschke:2005hs,Blaschke:2008}. (Below, we relate this to the widely-used WKB approach).  Within the quantum kinetic approach, the key quantity is the single-particle momentum distribution
function $f(\vec{k},t)$, which satisfies a non-Markovian quantum Vlasov equation including a source term for electron-positron pair production. We stress that $f(\vec{k},t)$ is physically meaningful as the distribution
function of real particles only at asymptotic times $t\rightarrow\pm\infty$, when the electric field vanishes. As demonstrated in \cite{Bloch:1999eu}, the field-current feedback due to Maxwell's equation can be neglected in the subcritical field strength regime, $E_0\lesssim 0.1E_\mathrm{cr}$. Thus, we need to solve just one integro-differential equation:
\begin{eqnarray}
  \label{eqn:qkeq}
  \frac{\mathrm{d}f(\vec{k},t)}{\mathrm{d}t}
  &=&\frac{1}{2}\frac{eE(t)\epsilon_\perp}{\omega^2(\vec{k},t)}
  \int_{-\infty}^{t}{\mathrm{d}t'\frac{eE(t')\epsilon_\perp}{\omega^2(\vec{k},t')}}
  \left[1-2f(\vec{k},t')\right] \nonumber \\ 
  &&\times\cos\left[2\int_{t'}^{t}{\mathrm{d}\tau\,\omega(\vec{k},\tau)}\right] \ .
\end{eqnarray}
Here $e$ is the electric charge; $\vec{k}=(\vec{k}_\perp,k_\parallel)$ is the canonical three-momentum; the kinetic momentum along the field is defined as $p_\parallel(t)=k_\parallel-eA(t)$; $\epsilon_\perp^2=m^2+\vec{k}_\perp^{\,2}$ is the transverse energy squared, and $\omega^2(\vec{k},t)=\epsilon_\perp^2+p_\parallel^2(t)$ characterizes the total energy squared. Note that Eq.~(\ref{eqn:qkeq}) is valid for Dirac particles (QED), however, there exists a very similar equation for scalar particles (sQED) which takes the statistics into account as well \cite{Kluger:1998bm, Smolyansky:1997fc,Schmidt:1998vi}.

The appearance of, and interplay between, a total of four scales -- electron mass $m$, applied electric field $E_0$,
laser frequency $\omega$, and total pulse length $\tau$ -- makes the electron-positron pair production process in such a laser field rather complicated \cite{Schutzhold:2008pz}, and suggests that the physics will not simply depend on the Keldysh parameter $\gamma\equiv m\omega/eE_0$. Schwinger pair production in a pulsed electric field Eq.~(\ref{eqn:elfield}), with zero carrier phase ($\phi=0$), has been studied using WKB \cite{Popov:2001}. In the nonperturbative regime, $\gamma\lesssim1$, the momentum spectrum is:
\begin{eqnarray}
  \label{eqn:popovmomdist}
  \frac{\mathrm{d}^3\mathcal{P}}{dk^3}\sim\exp\left(-\pi\frac{E_\mathrm{cr}}{E_0}
\left[1-\frac{1}{8}\tilde{\gamma}^2\right]
-\frac{1}{eE_0}\left[\tilde{\gamma}^2k_\parallel^2+\vec{k}_\perp^2\right]\right) 
\end{eqnarray}
with $\tilde\gamma^2=(1+1/\sigma^2)\gamma^2$. We will show that several interesting properties of the momentum distribution function for short pulses with many cycles per pulse are not captured by WKB.

We numerically integrate the quantum Vlasov equation Eq.~(\ref{eqn:qkeq}) to obtain the asymptotic distribution function $f(\vec{k},\infty)=\mathrm{d}^3\mathcal{P}/\mathrm{d}k^3$, for various values of the laser frequency $\omega$, choosing a temporal width $\tau=2\cdot10^{-4}\,\mathrm{eV^{-1}}$, which corresponds to a total pulse length of several times $10^{-19}\,\mathrm{s}$. This lies in the anticipated range of experimental parameters of future XFELs or may become realizable with higher harmonics or secondary-beam generation of optical lasers. We concentrate on the dependence on the longitudinal momentum $k_\parallel$, setting $\vec{k}_\perp=0$. For the more realistic case of additional spatial pulse inhomogeneities, we expect also nontrivial information encoded in all spatial momentum directions.

\begin{figure}[th]
  \centering
  \includegraphics[scale=0.8]{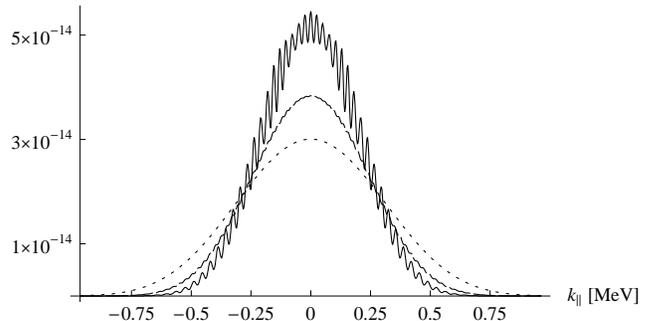}
  \caption{\label{fig:sigmadep} Asymptotic distribution function
    $f(\vec{k},\infty)$ for $\vec{k}_\perp=0$, $E_0=0.1E_\mathrm{cr}$ and
    $\phi=0$ when passing from $\sigma=3$ (dotted line) to $\sigma=4$ (dashed
    line) to $\sigma=5$ (solid line). Note that for $\sigma=3$ the
    canonical momentum value $k_\parallel=0$ corresponds to a kinetic momentum
    $p_\parallel(\infty)\approx75\,\mathrm{keV}$, whereas for $\sigma=5$
    the value $k_\parallel=0$ is equivalent to
    $p_\parallel(\infty)\approx0$.}
\end{figure}

The most dramatic new effect is that the momentum distribution shows distinctive oscillations, with an oscillation scale set by the laser frequency, as shown in Fig.~\ref{fig:sigmadep}. The distance between successive peaks in $f(\vec{k},\infty)$ is given by $\omega$. This oscillatory behavior becomes pronounced when $\sigma\gtrsim 4$, and the amplitude of the oscillations in the distribution function increases further as we increase the number
of cycles within the pulse beyond 4.

To understand the physical origin of these oscillations, we recall that for a spatially uniform but time dependent electric field, the pair production process can be viewed as a one dimensional quantum mechanical scattering
problem, with 'potential' given by $-\omega^2(\vec{k}, t)$ \cite{Brezin:1970xf,Popov:1972,Popov:1973az}. The number of produced pairs is related to the reflection coefficient for this over-the-barrier scattering problem. In fact, this scattering picture is completely equivalent to the quantum kinetic approach \cite{Dunne:2008,Dumlu:2009rr}, and so we can interpret these oscillations as being due to resonances in the scattering problem as the shape of the potential $-\omega^2(\vec{k}, t)$ changes with the momentum $\vec{k}$. These oscillations are missed when employing the WKB approximation to the scattering problem on which Eq.~(\ref{eqn:popovmomdist}) is based, but can be
seen clearly in an exact numerical integration of the scattering problem, or as we have shown here, in an exact numerical integration of the quantum Vlasov equation. This physical picture explains why the spacing between the peaks is the laser frequency, and also explains the sensitive dependence on the other shape parameters, such as $\sigma$.
\begin{figure}[t]
  \centering
  \includegraphics[scale=0.8]{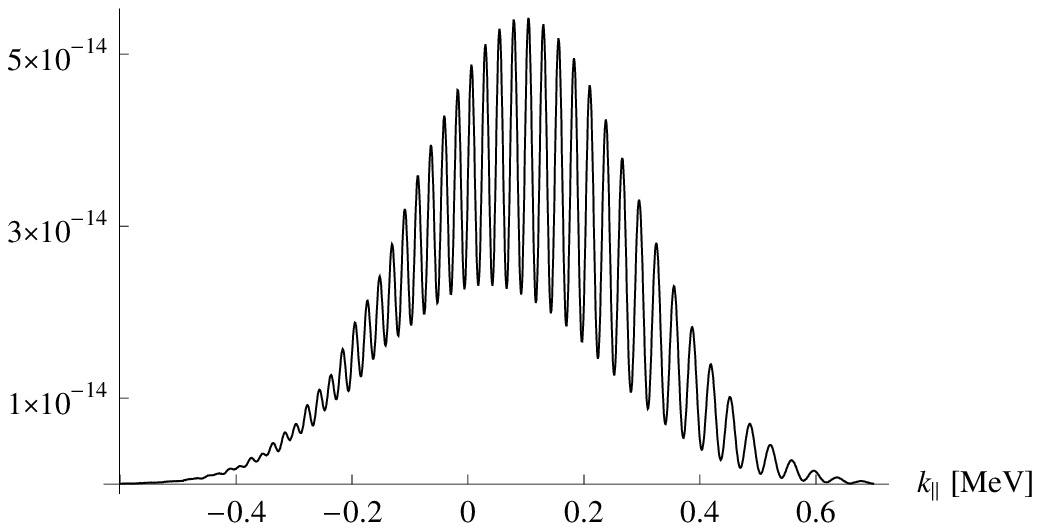}
  \caption{\label{fig:popovoscphipi4} Asymptotic distribution function
    $f(\vec{k},\infty)$ for $\vec{k_\perp}=0$ for $\sigma=5$,
    $E_0=0.1E_\mathrm{cr}$ and $\phi=-\pi/4$. The center of the distribution
    is shifted to $p_\parallel(\infty)\approx102\,\mathrm{keV}$.}
\end{figure}
\begin{figure}[t]
  \centering
  \includegraphics[scale=0.8]{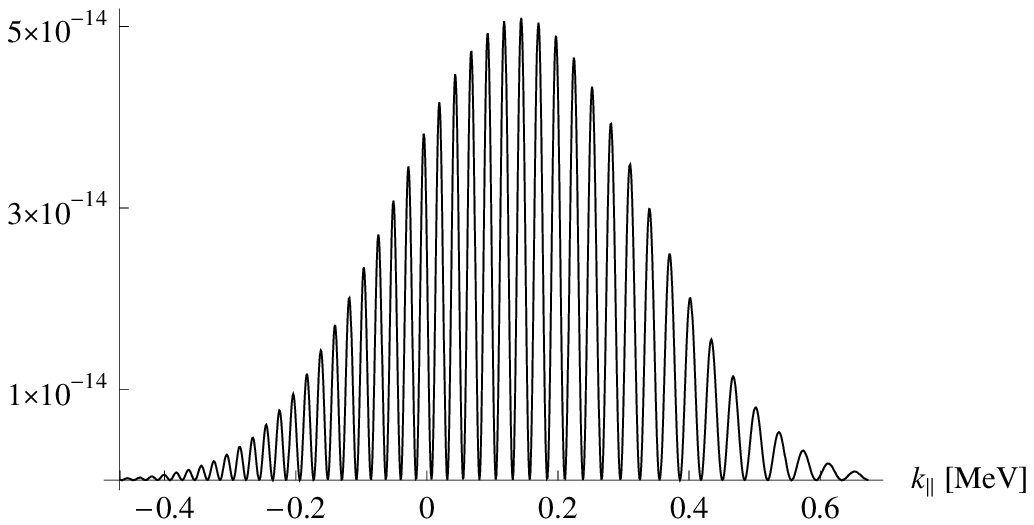}
  \caption{\label{fig:popovoscphipi2} Asymptotic distribution function
    $f(\vec{k},\infty)$ for $\vec{k_\perp}=0$ for $\sigma=5$,
    $E_0=0.1E_\mathrm{cr}$ and $\phi=-\pi/2$. The center of the distribution
    is shifted to $p_\parallel(\infty)\approx137\,\mathrm{keV}$.}
\end{figure}

In fact, there is an even more distinctive dependence on the carrier phase, $\phi$, upon which the form of the scattering potential $-\omega^2(\vec{k}, t)$ is extremely sensitive. The carrier phase dependence is difficult to discuss in the WKB approach, because a nonzero carrier phase breaks the $E(t)=E(-t)$ symmetry of the pulse shape, which in turn makes the imaginary time treatment of the WKB scattering problem significantly more complicated \cite{Popov:1972}.  But in the quantum kinetic approach, the carrier phase causes no computational problems;
it is just another parameter. We have found that the introduction of the carrier phase makes the oscillatory behavior in the longitudinal momentum distribution even more pronounced. This is shown in Figs.~\ref{fig:popovoscphipi4} and \ref{fig:popovoscphipi2}, where the momentum distribution function is plotted for $\phi=-\pi/4$ and $\phi=-\pi/2$. We see that for the same values of the other parameters, the oscillatory behavior becomes more distinct as the phase offset increases. The most distinctive momentum signature, however, is found for $\phi=-\pi/2$, when the electric
field is totally antisymmetric. In this case, the asymptotic distribution function $f(\vec{k},t)$ vanishes at the minima of the oscillations, as shown in Fig.~\ref{fig:popovoscphipi2}. This feature also has a direct analogue in the scattering picture: for an antisymmetric field, the gauge potential Eq.~\eqref{eqn:vecfield} is symmetric and so is the scattering potential well $-\omega^2(\vec{k}, t)$. In this case, perfect transmission is possible for certain resonance momenta, corresponding to zero reflection and thus zero pair production.  Also note that the center of the distribution shifts from $p_\parallel(\infty)=0$ to a non-zero value again. These carrier-phase effects
provide distinctive signatures, strongly  suggesting a new experimental strategy and probe in the search for Schwinger pair production.

These momentum signatures can also be understood in a quantum-mechanical double-slit picture, which has first been developed in the context of above-threshold ionization with few-cycle laser pulses \cite{Lindner:2005}: in this picture, the oscillations are fringes in the momentum spectrum that result from the interference of temporally separated pair creation events. The fringes are large for $\phi=-\pi/2$, since then the field strength has two peaks of equal size (though opposite sign) which act as two temporally separated slits. Moving the carrier phase away from $\phi=-\pi/2$ corresponds to gradually opening or closing the slits, resulting in a varying degree of which-way information and thus a varying contrast of the interference fringes. A quantitative consequence of this double-slit picture is that the width of the envelope of the oscillations in the distribution function is related to the
temporal width of the slits. The width of the envelope of oscillations thus also becomes a probe of the sub-cycle structure of the laser. 

\begin{figure}[t]
  \centering
  \includegraphics[scale=0.8]{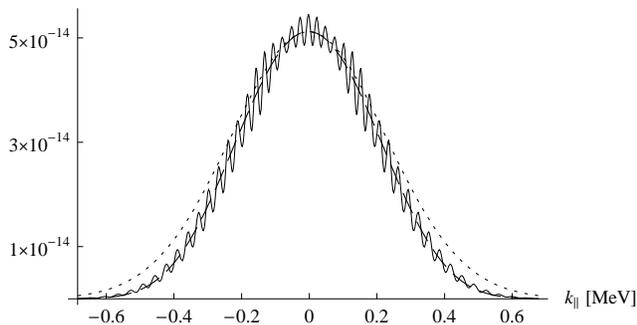}
  \caption{\label{fig:popovimp} Comparison of the asymptotic distribution
    function $f(\vec{k},\infty)$ for $\vec{k}_\perp=0$ (oscillating solid
    line) with the prediction of Eq.~(\ref{eqn:popovmomdist}) (dotted line)
    and the improved WKB approximation based on an expansion of
    Eq.~(\ref{eqn:contour}) (dashed line) for  $\sigma=5$, $E_0=0.1E_\mathrm{cr}$ and $\phi=0$.}
\end{figure}

To complete the physical picture, we consider the overall envelope of the longitudinal momentum distribution, again for $\phi=0$, averaging over the rapid oscillations.  When there are more than three cycles per pulse ($\sigma\gtrsim 3$), the peak of the momentum distribution is located near $p_\parallel(\infty)=0$, whereas for $\sigma\lesssim3$ the peak is shifted to a non-zero value. Furthermore, the Gaussian width of the employed WKB approximation Eq.~(\ref{eqn:popovmomdist}), which scales with $\sqrt{eE_0}/\tilde{\gamma}$, is obviously somewhat broader than the true distribution, as is shown in Fig.~\ref{fig:popovimp}.
We can quantify this discrepancy in the width, by extending the WKB result beyond the Gaussian approximation inherent in Eq.~(\ref{eqn:popovmomdist}). We use the results from \cite{Kim:2007pm}, where it has been shown that the WKB instanton action $S_{\vec{k}}$ in scalar QED can also be applied for spinor QED. Within this approach, $d^3\mathcal{P}/dk^3 \sim\exp (-2S_{\vec{k}})$, where the instanton action can be defined in the complex
$t-$plane as a contour integral:
\begin{equation}
  \label{eqn:contour}
  2S_{\vec{k}}=i\oint_{\Gamma}\sqrt{m^2+\vec{k}_\perp^2+[k_\parallel-eA(t)]^2}\,\mathrm{d}t \ ,
\end{equation}
with the path $\Gamma$ around the branch cut. After a change of variable, from $t$ to $T=-A(t)/E_0$, we can expand the instanton action $S_{\vec{k}}$ in powers of $(m/e E_0 \tau)^2$ and $(k_\parallel/e E_0 \tau)^2$. The result of this improved WKB expression to third order is plotted as the dashed line in Fig. \ref{fig:popovimp}, while the Gaussian WKB formula Eq.~(\ref{eqn:popovmomdist}) is shown as a dotted line.
We see that $\exp(-2S_{\vec{k}})$ fits very well the averaged envelope of the exact momentum distribution, while the Gaussian approximation Eq.~(\ref{eqn:popovmomdist}) is significantly broader. Neither WKB estimate sees the oscillatory structure of the momentum distribution. Furthermore, the oscillatory behaviour of the momentum distribution function is found also for scalar particles in the quantum kinetic framework of sQED, with the statistics playing a crucial role: the averaged envelope of the momentum distribution functions is identical, however, at momentum values where QED predicts a local maximum in the momentum distribution, sQED predicts a local minimum, and vice versa.

To conclude, we point out that the momentum distribution signatures described here, as well as their carrier-phase and oscillation-number dependence, provides for a new handle on the first detection of Schwinger pair production and its quantitative exploration, once laser field strengths have become sufficiently strong. In practice, these momentum signatures can distinguish Schwinger-produced pairs from possible background events, e.g., induced by residual-gas effects. As electron spectrometers used in laser-plasma acceleration experiments can reach a resolution of better than $1\%$, the whole spectrum including the oscillations will be directly accessible.

Analogously to strong-field ionization experiments \cite{Brabec:2000zz,Lindner:2005,Bohan:1998,Goulielmakis:2004}, these new pair-production signatures may also serve as sensitive probes of sub-cycle structure in ultra-short laser pulses. In addition to the positron yield, our results suggest a number of new observables such as the peak position and the width of the momentum distribution function and, most importantly, its potentially oscillatory structure. In particular, the characteristics of the oscillations can provide rather precise information about the carrier phase and the total pulse length. As the latter laser characteristics are difficult to control a priori in an absolute manner, these momentum signatures can serve as a tomograph of the laser pulse, providing for a unique means to verify and confirm design goals of future laser systems at highest intensities. 

{\bf Acknowledgements:} We thank M.~Kaluza, G.G.~Paulus and A.~Ringwald for helpful discussions. We acknowledge support from the DOC program of the Austrian Academy of Sciences and from the FWF doctoral college DK-W1203 (FH), from the US DOE grant DE-FG02-92ER40716 (GD), and from the DFG grant Gi328/5-1 and SFB-TR18 (HG).

\end{document}